\documentclass[twocolumn,twocolappendix]{aastex701}

\usepackage{amsmath}
\usepackage{bm}
\usepackage{CJK}
\usepackage{multirow}

\shorttitle{\textsc{Astrid}: Reduced Baryonic Suppression}
\shortauthors{Yang et al.}
\accepted{2026 July 19}
\published{2026 August 6}

\begin{document}

\title{Matter Clustering in \textsc{Astrid}: Reduced Baryonic Suppression from Realistic Black Hole Dynamics}

\author[orcid=0000-0001-6221-6024,sname='Yang',gname=Yanhui]{Yanhui Yang \begin{CJK}{UTF8}{gbsn}(杨焱辉)\end{CJK}}
\altaffiliation{Corresponding author}
\affiliation{Department of Physics and Astronomy, University of California, Riverside, 900 University Avenue, Riverside, CA 92521, USA}
\email[show]{yyang440@ucr.edu}

\author[orcid=0000-0001-5803-5490]{Simeon Bird}
\affiliation{Department of Physics and Astronomy, University of California, Riverside, 900 University Avenue, Riverside, CA 92521, USA}
\email[show]{sbird@ucr.edu}  

\author[0000-0002-8828-8461]{Yihao Zhou \begin{CJK}{UTF8}{gbsn}(周亦豪)\end{CJK}}
\affiliation{McWilliams Center for Cosmology, Department of Physics, Carnegie Mellon University, Pittsburgh, PA 15213, USA}
\email{yihaoz@andrew.cmu.edu}

\author[0000-0002-6462-5734]{Tiziana Di Matteo}
\affiliation{McWilliams Center for Cosmology, Department of Physics, Carnegie Mellon University, Pittsburgh, PA 15213, USA}
\email[show]{tiziana@phys.cmu.edu}

\author[0000-0003-0697-2583]{Rupert Croft}
\affiliation{McWilliams Center for Cosmology, Department of Physics, Carnegie Mellon University, Pittsburgh, PA 15213, USA}
\email{rcroft@cmu.edu}

\author[0000-0001-7899-7195]{Yueying Ni}
\affiliation{Michigan Institute for Data and AI in Society, University of Michigan, Ann Arbor, MI, 48109, USA}
\email{yueyingn@umich.edu}

\author[0000-0001-6627-2533]{Nianyi Chen}
\affiliation{School of Natural Sciences, Institute for Advanced Study, Princeton, NJ 08540, USA}
\email{nianyi.chen7@gmail.com}

\begin{abstract}

Baryonic feedback from active galactic nuclei (AGN) is often invoked as a major source of suppression in the matter power spectrum, with implications for precision cosmology and the $S_8$ tension. We present \textsc{Astrid-DMO}, the dark matter-only counterpart to the large-volume \textsc{Astrid} hydrodynamical simulation, and measure baryonic effects through $P_{\rm hydro}(k)/P_{\rm DMO}(k)$. We find no significant suppression at $z=0$ and mild suppression at $z=0.2$, weaker than in other state-of-the-art simulations. Using controlled small-volume runs, we identify a key driver of this discrepancy: the treatment of black hole (BH) dynamics. The widely used BH repositioning scheme artificially enhances BH mergers and boosts kinetic AGN feedback (e.g., by a factor of $2$ at $z=1.5$), leading to overly strong suppression. By contrast, a more physical dynamical friction model reduces feedback efficiency and weakens clustering suppression. Consequently, reconciling large-scale structure measurements with cosmic microwave background (CMB)-inferred $\Lambda$CDM cosmology, while matching observed halo gas fractions, becomes more challenging. Although strengthening AGN feedback can increase suppression, in our model this induces tensions with the observed galaxy stellar mass and AGN luminosity functions. These results sharpen the need for novel mechanisms that can efficiently eject gas from halos without compromising other galaxy properties.
\end{abstract}


\section{Introduction\label{sec:intro}} 

Interpreting measurements of cosmic large-scale structure \citep[e.g.,][]{Ivezic2019,Aghamousa2016,Aihara2018,Laureijs2011,Akeson2019,Gong2019} requires a robust treatment of baryonic physics to fully exploit their constraining power. This is particularly relevant to the $S_8$ tension \citep[for an introduction, see][]{Hildebrandt2017,Abbott2018,Heymans2021}, where baryonic suppression has been proposed as a potential solution \citep{McCarthy2018,Amon2022}, though disagreements remain in the community. For instance, the recent reanalysis of the Hyper Suprime-Cam (HSC) Y3 data \citep{Choppin2025} models baryonic effects and derives a cosmology consistent with cosmic microwave background (CMB) measurements \citep{Qu2026,Louis2025}. In contrast, the Dark Energy Survey (DES) Y6 analysis \citep{DES2026} adopts conservative scale cuts that exclude data potentially sensitive to baryonic uncertainties, and continues to find a $2.6\sigma$ $S_8$ discrepancy. Several studies argue that baryonic physics may not fully resolve the tension, as the required suppression is too strong to be consistent with observed properties of galaxy groups and clusters, such as the gas/baryon fraction \citep[e.g.,][]{McCarthy2024,Salcido2025}, implying that new physics (e.g., new dark sector models) may be required.

Active galactic nucleus (AGN) feedback is the dominant source of baryonic suppression of matter clustering on Mpc scales \citep{vanDaalen2011,Chisari2018}. In modern cosmological hydrodynamical simulations \citep[e.g., the IllustrisTNG simulations;][]{Springel2018}, kinetic AGN feedback, associated with high-mass black holes (BHs) accreting at low rates, is implemented to efficiently remove gas from halos without excessive galaxy sizes or X-ray luminosities associated with older models \citep[e.g.,][]{Snyder2015,Choi2015}. Recent large-volume simulations with AGN feedback implemented have shown considerable baryonic suppression of the matter power spectrum (tens of percent at $k\sim 10\,h\,$Mpc$^{-1}$) \citep[e.g.,][selected simulations are shown in Fig.~\ref{fig:pk_kfeedback}]{Springel2018,Chisari2018,Schaye2023,Hellwing2016}.

The \textsc{Astrid} simulation \citep{Zhou2026}, one of the largest high-resolution cosmological hydrodynamical simulations to date, also includes a kinetic AGN feedback model. In addition, it implements a relatively realistic treatment of BH dynamics by including a subgrid dynamical friction (DF) model \citep{Chen2022}, which allows BHs to orbit within their host halos and merge only when they are sufficiently close and gravitationally bound \citep[see also][]{Hirschmann2014,Tremmel2015,Tremmel2017}. In contrast, most existing simulations artificially pin (or reposition) BHs to the centers (i.e., gravitational potential minima) of their host galaxies \citep[e.g.,][]{Booth2009,Springel2018}.\footnote{This technique is used to avoid numerical dynamical heating of particles caused by insufficient mass resolution.} This treatment can lead to unphysically prompt BH mergers during early close passages of their host halos. \citet{Chen2022} found that, in simulations with repositioning, BHs of small (sub)halos tend to merge into massive BHs in larger halos,\footnote{In Fig.~7 of \cite{Chen2022}, there are many halos that have lost their BHs.} resulting in artificially enhanced growth of high-mass BHs \citep[see Fig.~9 of][]{Chen2022}. This effect is also observed in this work: as shown in Figure~\ref{fig:BH_halo_Ekf}, the number of BHs with $M_\text{BH} < 10^6\,\mathrm{M}_\odot/h$ is significantly lower in the repositioning simulations. Whether this effect can alter AGN feedback and thus matter clustering has not yet been carefully investigated.

In this work, we present \textsc{Astrid-DMO}, the dark matter-only (DMO) counterpart simulation to \textsc{Astrid}, and examine baryonic effects on the matter power spectrum in \textsc{Astrid}. To compare DF and repositioning in their effects on baryonic suppression, we perform and analyze a set of smaller-volume simulations with varying BH dynamics and feedback configurations. We also present and discuss selected observables affected by these variations.

\section{Simulations\label{sec:sims}} 

\subsection{\textsc{Astrid} and \textsc{Astrid-DMO}}

The \textsc{Astrid} simulation evolves $2\times 5500^3$ particles in a $250$ cMpc$/h$ box, with a gravitational softening length of $\epsilon_\mathrm{g} = 1.5\,\mathrm{ckpc}/h$ \citep{Bird2022,Ni2022,Ni2025,Zhou2026}. Importantly, \textsc{Astrid} includes a realistic dynamical friction model described in \cite{Chen2022} and a model for kinetic-mode black hole feedback \citep{Weinberger2017}. This BH treatment allows us to track the orbital evolution of BHs and to impose a physical criterion for BH mergers that requires both sufficient proximity and gravitational binding \citep{Bellovary2011,Tremmel2017}. For the criterion adopted in \textsc{Astrid}, see Equation~(1) in \cite{Zhou2026}.

\textsc{Astrid} employs a two-mode AGN feedback model, with a thermal mode at high accretion rates and a kinetic mode at low accretion rates. Following \cite{Weinberger2017}, the mode-switching criterion is based on both the BH mass and the Eddington ratio, $f_\mathrm{Edd} = \dot{M}_\mathrm{BH}/\dot{M}_\mathrm{Edd}$, where $\dot{M}_\mathrm{BH}$ ($\dot{M}_\mathrm{Edd}$) is the BH (Eddington) accretion rate. The threshold for activating kinetic feedback is defined as:
\begin{equation}
\chi_\mathrm{thr} = \mathrm{min} \left[0.002\left(\frac{M_\mathrm{BH}}{M_\mathrm{pivot}}\right)^2, 0.05\right],
\end{equation}
where the pivot mass $M_\mathrm{pivot}$ is set to $5\times 10^8\,\mathrm{M}_\odot/h$.
When $f_\mathrm{Edd} < \chi_\mathrm{thr}$, AGN feedback switches to the kinetic mode, in which energy is injected in the form of momentum kicks to surrounding gas particles. In the thermal (quasar) mode, the feedback energy is deposited as thermal energy into nearby gas. 

We note that this kinetic-mode activation criterion is more stringent than simulations such as IllustrisTNG \citep{Springel2018}, which also contributes to weakening baryonic suppression in \textsc{Astrid} by reducing the number of BHs that can enter the kinetic mode \citep{Ni2023}.

We perform \textsc{Astrid-DMO}, with the same cosmological parameters, volume and initial phases as \textsc{Astrid}, except that it contains $5500^3$ dark matter particles and no baryons. \textsc{Astrid-DMO} also evolves from $z=99$ to $0$, and has particle snapshots and halo catalogs saved at the same redshifts as \textsc{Astrid}, with $z < 3$ halo catalogs saved at a finer redshift cadence for merger tree generation. We use the Friends-of-Friends (FoF) algorithm \citep{Davis1985} to identify particle groups and \textsc{subfind} \citep{Springel2001, Springel2010} to identify halos and subhalos.

\subsection{Small-volume simulations}

We investigate how the matter power spectrum is affected by different BH model choices with an additional set of small-volume (SV) simulations.
Each hydrodynamical simulation includes $2\times 640^3$ particles initially in a volume of (50$\,$cMpc$/h)^3$, with particle masses of $M_\mathrm{DM} = 3.43\times 10^7\,\mathrm{M}_\odot/h$ and $M_\mathrm{gas} = 6.43\times 10^6\,\mathrm{M}_\odot/h$, and a softening length of $\epsilon_\mathrm{g}^\mathrm{SV}=2.6\,\mathrm{ckpc}/h$ ($2\times$ lower resolution than \textsc{Astrid}).
The simulations are summarized in Table~\ref{tab:smallsims}. \textsc{SV-DF} uses the same BH dynamics model (dynamical friction) and AGN feedback model (fiducial) as \textsc{Astrid}. In addition, to verify that the effects of BH dynamics are generalizable to different feedback strengths, we run two more simulations with stronger AGN feedback, \textsc{SV-DF-S} and \textsc{SV-Repos-S}. The stronger feedback model is achieved by loosening the kinetic mode activation criterion and increasing the kinetic feedback efficiency \citep[for the definition of the efficiency $\epsilon_\mathrm{f,kin}$, see][]{Zhou2026}. Specifically, we change the activation threshold by setting the pivot mass to 1/5 of the fiducial value, i.e., $M_\mathrm{pivot} = 10^8\,\mathrm{M}_\odot/h$, and double the cap from 0.05 to 0.1, such that BHs can enter the kinetic mode more easily (we will see in Fig.~\ref{fig:KE_density_BH} that even low-mass BHs can release considerable energy). The kinetic feedback efficiency is enhanced by a factor of 4. All other cosmological and subgrid model parameters are the same as in \textsc{Astrid}.
We have also run \textsc{SV-DMO}, a DMO counterpart of these small simulations with $640^3$ dark matter particles in a box of the same size. All simulations are run with \textsc{MP-Gadget} \citep{Feng2018} throughout this study.

In each hydrodynamical simulation, we track BH activity at every time step, recording instantaneous accretion rates, energy release, and merger events. This high-cadence logging generates a continuous evolutionary history, capturing transient dynamical events and feedback phases that would otherwise be missed between snapshots.

\begin{table}[h!]
\centering
\caption{Small-volume simulations performed to study the effect of BH dynamics models on the matter power spectrum suppression. `Repositioning' refers to the BH repositioning method. AGN feedback models are varied to confirm the generalizability of the effects of BH dynamics.}
\begin{tabular}{lcc}
\hline
\hline
Simulation & BH dynamics & Feedback model \\
\hline
\textsc{SV-DF} & Dynamical friction & Fiducial \\
\textsc{SV-Repos} & Repositioning & Fiducial \\
\textsc{SV-DF-S} & Dynamical friction & Stronger feedback \\
\textsc{SV-Repos-S} & Repositioning & Stronger feedback \\
\hline
\end{tabular}
\label{tab:smallsims}
\end{table}

\section{Results\label{sec:results}} 

\subsection{Baryonic suppression of matter clustering\label{sec:clustering}}

\begin{figure*}[ht!]
     \centering
     \includegraphics[width=\textwidth]{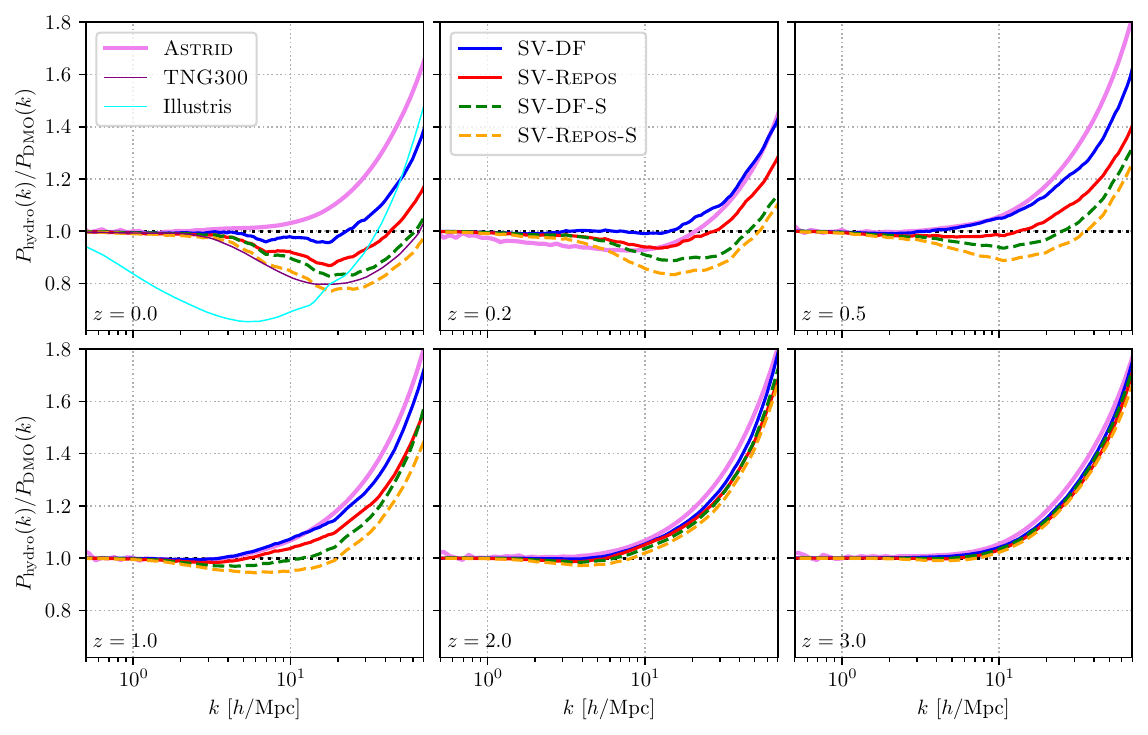}
     \caption{Matter power spectra measured from the \textsc{Astrid} simulation and the small simulations (listed in Table~\ref{tab:smallsims}) normalized by their corresponding DMO counterparts, at $z=0, 0.2, 0.5, 1, 2$ and $3$ in six panels respectively. For comparison, we also plot the $P(k)$ suppression for the Illustris and TNG300 simulations at $z=0$. Different simulations are color coded. Solid curves correspond to simulations with the fiducial feedback model, while dashed curves represent simulations with the stronger feedback model. The horizontal dotted line marks unity.\label{fig:pk_kfeedback}}
\end{figure*}

We plot the ratio of the matter power spectrum between the hydrodynamical and DMO simulations, $P_\mathrm{hydro}(k)/P_\mathrm{DMO}(k)$, for both the \textsc{Astrid} and SV simulations in Figure~\ref{fig:pk_kfeedback}. In all simulations, the ratio is close to unity at large scales ($k\lesssim 1\,h\,$Mpc$^{-1}$) where baryonic effects are negligible, and much greater than unity at the smallest scales due to the cooling and condensation of baryons in the centers of halos.

In \textsc{Astrid} (violet), no suppression is seen prior to $z=0.5$. At $z=0.2$, mild suppression extends to $k\sim 1\,h\,\mathrm{Mpc}^{-1}$, before the power spectrum ratio rises back to above unity at $z=0$; this behavior may reflect gas reaccretion by sufficiently massive halos and/or continued adiabatic contraction of dark matter.\footnote{We would not expect this behavior in simulations with smaller volumes and stronger feedback.} By contrast, Illustris and TNG300 show significantly suppressed matter power spectra.\footnote{We note that TNG300 has a mass resolution similar to that of \textsc{SV-Repos} and is nearly identical to \textsc{SV-Repos} in its kinetic feedback activation criterion and efficiency, except that it adopts a lower $M_\mathrm{pivot}$ of $10^{8}\,\mathrm{M}\odot = 0.6774\times 10^{8}\,\mathrm{M}\odot/h$, making kinetic feedback easier to activate.} \textsc{Astrid}'s SV counterpart, \textsc{SV-DF} (blue), differs because of its smaller box size and lower resolution. Its reduced small-scale power is due to the lower resolution, which increases BH mergers through the resolution (softening length)-dependent merging criterion. Accordingly, DF and repositioning are less distinguishable at lower resolution, as verified in Appendix~\ref{app:pk_supplementary}. Because \textsc{SV-DF} does not form halos as massive as in \textsc{Astrid}, it does not show the extended suppression or the subsequent recovery. Although \textsc{SV-DF} is not fully converged, it is sufficient for examining the relative effects of the feedback variations.

The SV runs show different baryonic effects on the matter power spectrum. \textsc{SV-Repos} (red) shows a stronger suppression than \textsc{SV-DF} (blue) at all redshifts; e.g., at $z=0$ the power ratio of \textsc{SV-Repos} is lower than that of \textsc{SV-DF} by $\sim 10\%$ around $k= 20\,h\,\mathrm{Mpc}^{-1}$ (we expect the discrepancy to be $\gtrsim 20\%$ for \textsc{Astrid}'s resolution). This implies that the repositioning method leads to more efficient AGN feedback. The trend holds for the strong-feedback models (dashed), where \textsc{SV-Repos-S} (orange) shows stronger suppression than \textsc{SV-DF-S} (green), though the discrepancy is less significant than in the fiducial feedback runs (the relative difference is alleviated by the increased number of kinetic-mode BHs, see Appendix~\ref{app:BHscatter}). 

\begin{figure*}[ht!]
     \centering
     \includegraphics[width=\textwidth]{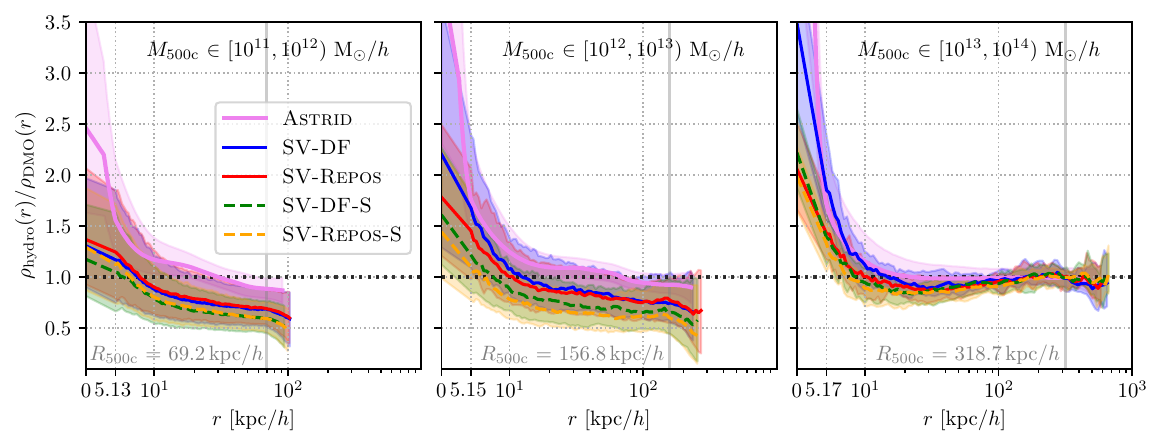}
     \caption{Halo density profile ratios of halos in the hydrodynamical simulations to their matched DMO counterparts, at $z=0$. The color and line styles are the same as in Figure~\ref{fig:pk_kfeedback}. Unity is marked by a horizontal dotted line. Left, middle and right panels show ratios for halos in the mass ranges of $M_\mathrm{500c} =$ $10^{11}$--$10^{12}\,\mathrm{M}_\odot/h$, $10^{12}$--$10^{13}\,\mathrm{M}_\odot/h$ and $10^{13}$--$10^{14}\,\mathrm{M}_\odot/h$, respectively. Shaded areas represent the 16th to 84th percentile scatter among the matched halo population. Ratios at radii within a certain scale ($\sim 2\epsilon_\mathrm{g}^\mathrm{SV}$), $r_\mathrm{lin}$, are plotted on a linear scale to cover the halo centers, while ratios at larger radii are plotted with a logarithmic $x$-axis. $r_\mathrm{lin} = 5.13$, $5.15$ and $5.17\,\mathrm{kpc}/h$ for the three mass bins, respectively. Median $R_\mathrm{500c}$ value of the DMO halos (measured from \textsc{SV-DMO}) in each mass bin is shown by a vertical grey line in the corresponding panel.
     \label{fig:densityprofile_kfeedback}}
\end{figure*}

To compare the baryonic effects of different models in more detail, we match halos in the hydrodynamical simulations to their DMO counterparts at $z=0$ and plot the ratios of their density profiles, $\rho_\mathrm{hydro}(r)/\rho_\mathrm{DMO}(r)$, in Fig.~\ref{fig:densityprofile_kfeedback} (see Appendix~\ref{app:halomatching} for details on the matching procedure). $\rho_\mathrm{hydro}(r)$ and $\rho_\mathrm{DMO}(r)$ are the one-dimensional total matter density profiles of the matched halos in the hydrodynamical and DMO simulations, respectively. Halo masses are expressed as $M_\mathrm{500c}$, the mass enclosed within the radius centered on the potential minimum where the mean density is 500 times the critical density of the Universe.

The density profile ratios show systematic differences among the models. Comparing \textsc{SV-DF} (blue) with \textsc{SV-Repos} (red), we see that the repositioning method leads to lower matter density at $r\lesssim 100\,\mathrm{kpc}/h$ in the intermediate- (middle) and high-mass (right) bins, which is consistent with the stronger suppression of the matter power spectrum in \textsc{SV-Repos} than in \textsc{SV-DF} observed in Fig.~\ref{fig:pk_kfeedback}. In the low-mass bin (left panel), the difference between \textsc{SV-DF} and \textsc{SV-Repos} is less significant, because there are only small numbers of kinetic-mode BHs in these halos.
In the strong AGN feedback runs (dashed), while the overall density profiles shift downwards, we observe a similar trend as in the fiducial feedback runs, though the relative differences are smaller. \textsc{Astrid}, as expected from the matter power spectrum analysis (Fig.~\ref{fig:pk_kfeedback}), shows a higher $\rho_\mathrm{hydro}(r)/\rho_\mathrm{DMO}(r)$, especially in the inner regions of the halo, than the SV runs.

\subsection{Kinetic feedback energy\label{sec:feedback}}

\begin{figure*}[ht!]
     \centering
     \includegraphics[width=\textwidth]{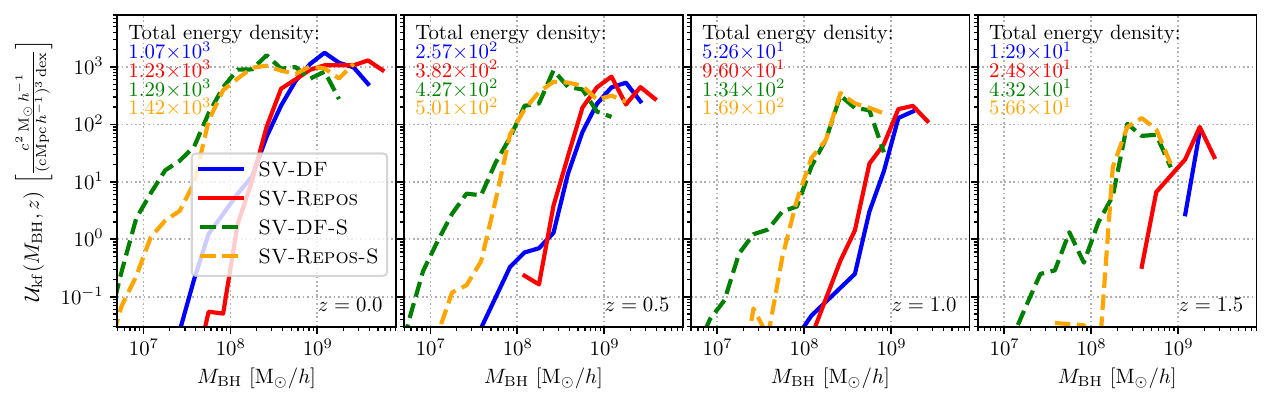}
     \caption{Cumulative released kinetic feedback energy density distribution as a function of BH mass. The color and line styles are the same as in Figure~\ref{fig:pk_kfeedback}. Panels from left to right show results at $z = 0$, $0.5$, $1$, and $1.5$. Total released kinetic feedback energy [i.e., $U_\mathrm{kf} (\infty , z)$] is annotated for each SV simulation in the corresponding color. \label{fig:KE_density_BH}}
\end{figure*}

In this section, we analyze how the BH dynamics treatment affects kinetic AGN feedback. We define $U_\mathrm{kf}(M_\mathrm{BH}, z)$, the density of the cumulative kinetic feedback energy released from BHs with masses $< M_\mathrm{BH}$ by redshift $z$. The differential kinetic feedback energy density per logarithmic BH-mass interval is then
\begin{equation}
\mathcal{U}_\mathrm{kf}(M_\mathrm{BH}, z) = \frac{\partial\,U_\mathrm{kf}(M_\mathrm{BH}, z)}{\partial \log (M_\mathrm{BH}\,[\mathrm{M}_\odot/h])}.
\end{equation}
Throughout this Letter, energy density is in units of $h^{2}\,\mathrm{M}_\odot\,c^2\,\mathrm{cMpc}^{-3}$.
At a given redshift, the kinetic energy assigned to each BH includes contributions from all BHs that have merged into it (i.e., its progenitors).

Figure~\ref{fig:KE_density_BH} shows $\mathcal{U}_\mathrm{kf}(M_\mathrm{BH}, z)$ at $z=0$, $0.5$, $1$ and $1.5$ in four panels for the SV simulations. 
\textsc{SV-Repos} (red) releases more kinetic feedback energy than \textsc{SV-DF} (blue), with relative differences more pronounced at early times. For instance, at $z=1.5$, the total kinetic feedback energy released in \textsc{SV-Repos} is about twice that in \textsc{SV-DF}. We note that early kinetic feedback removes gas very efficiently despite the still modest cumulative energy injection (we have explicitly verified this in Fig.~\ref{fig:pk_Skz1}). Figure~\ref{fig:KE_density_BH} also shows that the most massive BHs in \textsc{SV-Repos} have higher masses than in \textsc{SV-DF}.
The excess kinetic energy release in the repositioning method is driven by more BHs entering the kinetic mode, ultimately from the spurious mergers caused by repositioning (see also Appendix~\ref{app:BHscatter}). 

Though the discrepancy decreases with time, \textsc{SV-Repos} still releases about 15\% more kinetic feedback energy than \textsc{SV-DF} at $z=0$. We notice that \textsc{SV-DF} has more energy released at the low-mass end than \textsc{SV-Repos}, which is because the latter has significantly fewer low-mass BHs. However, the contribution of low-mass BHs to the total kinetic feedback energy is negligible.

In the strong-feedback runs (dashed), the earlier kinetic-mode activation and increased kinetic feedback efficiency lead to more efficient gas expulsion at early times and suppress the growth of BHs; hence, we see lower maximum BH masses than in the fiducial feedback runs. Nevertheless, we still find the repositioning run has more kinetic feedback energy injected than the DF run, despite a smaller fractional difference than in the fiducial feedback case [e.g., $U_\mathrm{kf}^\textsc{SV-Repos-S} (\infty , 0)$ is 10\% greater than $U_\mathrm{kf}^\textsc{SV-DF-S} (\infty , 0)$, while $U_\mathrm{kf}^\textsc{SV-Repos} (\infty , 0)$ is larger than $U_\mathrm{kf}^\textsc{SV-DF} (\infty , 0)$ by 15\%].

Overall, this analysis indicates that the repositioning method artificially boosts kinetic AGN feedback (consistent with the matter clustering results shown in Section~\ref{sec:clustering}). 

\subsection{AGN X-ray luminosity function}

\begin{figure}[ht!]
     \centering
     \includegraphics[width=\linewidth]{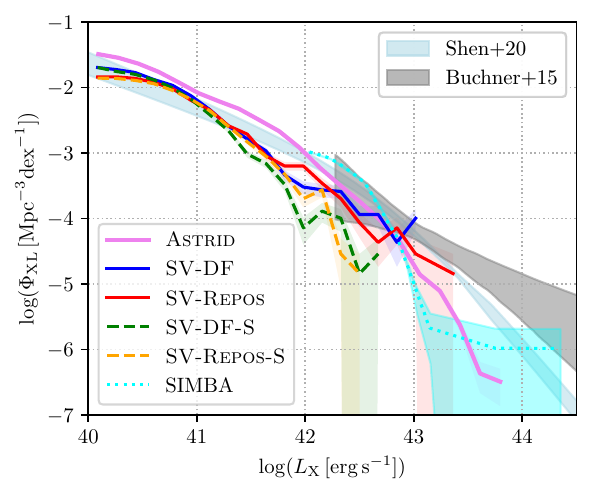}
     \caption{AGN hard X-ray (2--10$\,$keV) luminosity functions at $z=0$. The line styles and colors are the same as in Fig.~\ref{fig:pk_kfeedback}. The SIMBA luminosity function \citep[data from][]{Habouzit2022} is also plotted for comparison. Shaded regions show observational constraints compiled by \cite{Shen2020} (light blue) and \cite{Buchner2015} ($z=0.1$, gray).}
     \label{fig:BH_XLF}
\end{figure}

We present the AGN hard X-ray (2--10$\,$keV) luminosity functions at $z=0$ for the \textsc{Astrid} and SV simulations in Fig.~\ref{fig:BH_XLF}. The BH X-ray luminosity ($L_\mathrm{X}$) is calculated following the method of \cite{Hopkins2007}. We show results assuming a fixed radiative efficiency of $\eta=0.1$ \citep{Shakura1973}. For alternative choices of $\eta$ and related discussion, please refer to \cite{Zhou2026}. We remind the reader that the SV simulations are not fully converged and thus used only to compare the relative effects of BH dynamics and feedback models, rather than for direct comparison with observations.

In this figure, one observes that \textsc{Astrid} agrees well with the observational constraints except at the bright end, where it underpredicts the number density of luminous AGN. Within the SV runs, the strong-feedback simulations yield significantly fewer luminous AGN than the fiducial runs. This trend is expected: in the strong-feedback regime, more gas is expelled from BH neighborhoods (see Fig.~\ref{fig:densityprofile_kfeedback}) and the BHs themselves are less massive (see Fig.~\ref{fig:BH_halo_Ekf}). However, changing the BH dynamics model (DF versus repositioning) does not yield a clear separation in the $z=0$ luminosity functions. A likely explanation is compensating evolution: although \textsc{SV-Repos} has more high-mass BHs at early times, its more efficient kinetic feedback removes more gas and thus suppresses later accretion, bringing its luminosity function at $z=0$ close to that of \textsc{SV-DF} (consistent with the behavior of $U_\mathrm{kf}$ discussed in Sec.~\ref{sec:feedback}). We note that the luminosity functions of the SV simulations are relatively noisy at the bright end because of the small volume; future work with larger volumes will further test the effects of BH dynamics on the luminosity function.

For reference, we also plot the SIMBA luminosity function; it likewise shows a bright-end deficit, although this comparison is not strictly one-to-one because the underlying BH models differ from those in \textsc{Astrid}.

We provide results for additional observables in appendices, including the galaxy stellar mass function (GSMF) in Appendix~\ref{app:GSMF} and halo gas fractions in Appendix~\ref{app:gasfraction}. The former confirms that more efficient AGN feedback (either from BH dynamics treatment or feedback model) can suppress star formation in massive galaxies, while the latter shows that \textsc{Astrid} produces higher gas fractions than other simulations, which is consistent with its reduced baryonic suppression of matter clustering.

\section{Conclusions and discussion\label{sec:discussion}} 

In this study, we have presented the baryonic matter power spectrum suppression of \textsc{Astrid} compared to the new \textsc{Astrid-DMO} simulation. In Fig.~\ref{fig:pk_kfeedback}, while there is mild suppression at $z=0.2$, we do not observe significant suppression at $z=0$. A set of small-volume simulations has been performed to compare kinetic AGN feedback and baryonic suppression between the repositioning and dynamical friction methods for BH dynamics modeling. These simulations identify the improved and more physical BH dynamics model as a key reason for our reduced suppression compared to other simulations. Compared to the widely used repositioning method, the DF model treats BH merging more physically and yields less kinetic AGN feedback, leading to significantly weaker suppression of matter clustering, under the assumption that higher kinetic feedback energy corresponds to stronger suppression. This suggests that existing simulations exhibiting strong matter clustering suppression could show weaker suppression if feedback artifacts associated with BH repositioning were removed.

The reduced $P(k)$ suppression makes it more difficult to reconcile the observed structure growth with the $\Lambda$CDM model by invoking baryonic feedback. In line with this, \textsc{Astrid} produces significantly higher gas fractions in halos than other simulations and observational estimates (Appendix~\ref{app:gasfraction}). Although simply increasing feedback strength can expel gas more efficiently and produce stronger suppression of the matter power spectrum, it also introduces undesired side effects in our model, including fewer luminous AGN at $z=0$ (Fig.~\ref{fig:BH_XLF}) and overquenched star formation in massive galaxies (Appendix~\ref{app:GSMF}).

Our findings strengthen the need for novel mechanisms capable of expelling baryons from halos without excessively altering other galaxy properties. Possible solutions include revising the BH seeding prescription or incorporating other feedback channels, such as jet feedback \citep{Dave2019} or variants of the radio-mode feedback used in Illustris \citep{Vogelsberger2014}. We advocate exploring these possibilities with realistically modeled BH dynamics; otherwise, unphysical feedback artifacts might obscure genuine signatures of new physics, such as nontrivial dark matter.
An additional question for future work is whether different BH dynamics methods affect how efficiently AGN feedback removes gas from halos, independently of their different BH merging histories, given that repositioning can produce burstier AGN feedback \citep{Wurster2013,Tremmel2017}. Answering this will require more controlled experiments.

\section*{Data Availability}
The \textsc{Astrid-DMO} and SV simulation data will be shared upon reasonable request to the corresponding author. The analysis code used in this work has been made publicly available at \cite{CodeYang2026}.

\begin{acknowledgments}
S.B. and Y.Y. were supported in part by grant 63667 from the John Templeton Foundation. The opinions expressed in this publication are those of the author(s) and do not necessarily reflect the views of the John Templeton Foundation. S.B. and Y.Y. were supported by NSF AST-2509639. Y.Z. and T.D.M. acknowledge the support from the NASA FINESST grant 80NSSC25K0318. T.D.M. acknowledges funding from NASA ATP 80NSSC20K0519, NSF PHY-2020295, NASA ATP NNX17AK56G, NASA Theory grant 80NSSC22K072, and NASA ATP 80NSSC18K101. Y.N. acknowledges support from the ITC Postdoctoral Fellowship. N.C. acknowledges support from the Schmidt Futures Fund and MPA Postdoctoral Fellowship. ASTRID was run on the Frontera facility at the Texas Advanced Computing Center. The authors used ChatGPT \citep{openai2026} to assist with improving the clarity and language of the manuscript. The authors take full responsibility for the content of this publication.
\end{acknowledgments}

\appendix
\numberwithin{figure}{section}
\renewcommand{\thefigure}{\thesection\arabic{figure}}
\numberwithin{table}{section}
\renewcommand{\thetable}{\thesection\arabic{table}}

\section{BH--halo mass relation and feedback energy\label{app:BHscatter}}

\begin{figure*}[ht!]
     \centering
     \includegraphics[width=\textwidth]{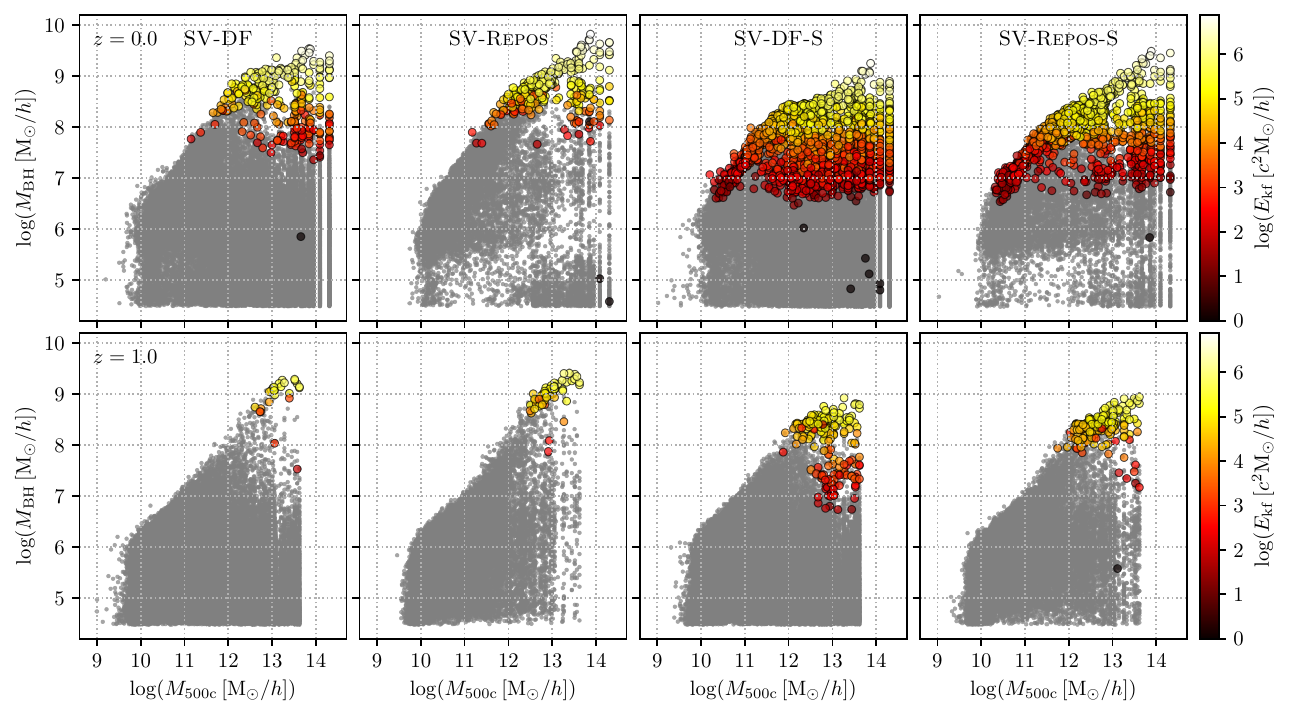}
     \caption{Scatter plot of BH mass ($M_\mathrm{BH}$) versus halo mass ($M_\mathrm{500c}$), colored by the cumulative released kinetic feedback energy $E_\mathrm{kf}$. The first and second rows show the relation at $z=0$ and $z=1$, respectively. The four columns correspond to the four different SV simulations. Color circles represent BHs that have released kinetic feedback energy, while gray points represent BHs that have not released kinetic feedback energy.}
     \label{fig:BH_halo_Ekf}
\end{figure*}

Figure~\ref{fig:BH_halo_Ekf} plots masses of individual BHs versus their host halo masses at $z=0$ (top row) and $z=1$ (bottom row) for the four SV simulations. Looking at the first two columns, we see that the repositioning method (column (2)) leads to more massive BHs at the high-mass end than the DF method (column (1)). In \textsc{SV-Repos}, we also find that there are fewer low-mass BHs in high-mass halos than in \textsc{SV-DF}, which is because the repositioning method makes central BHs of low-mass subhalos more likely to be swallowed by the central BH of the main halo. For example, at $z=1$ and $M_\mathrm{500c} \gtrsim 10^{13}\,\mathrm{M}_\odot/h$, \textsc{SV-Repos} has visibly fewer BHs with $M_\mathrm{BH} \lesssim 10^6\,\mathrm{M}_\odot/h$ than \textsc{SV-DF}. Compared to the fiducial runs (columns (1) and (2)), the strong-feedback runs (columns (3) and (4)) show lower maximum BH masses and more extended distributions of BHs that have released kinetic feedback energy, which is expected given the looser kinetic-mode activation criterion. We also see \textsc{SV-Repos-S} differs from \textsc{SV-DF-S} in a similar way as \textsc{SV-Repos} differs from \textsc{SV-DF}, e.g., more massive BHs at the high-mass end.

\section{$P(k)$ suppression in supplementary runs\label{app:pk_supplementary}}

\begin{figure}[ht!]
     \centering
     \includegraphics[width=\linewidth]{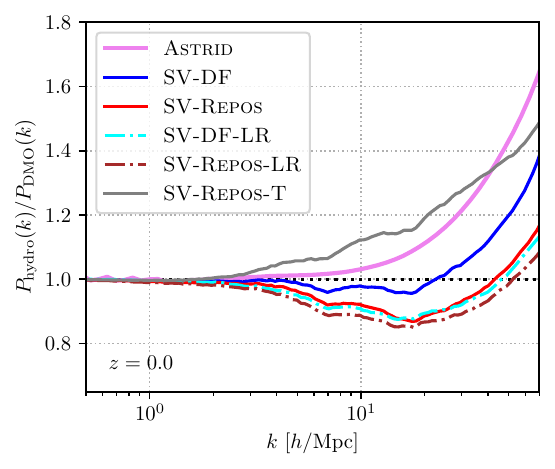}
     \caption{Comparison of $P(k)$ suppression among the simulations with different resolutions and feedback types at $z=0$. The color and line styles are the same as in Fig.~\ref{fig:pk_kfeedback}, with the addition of \textsc{SV-DF-LR} (dashed--dotted cyan), \textsc{SV-Repos-LR} (dashed--dotted brown) and \textsc{SV-Repos-T} (solid gray). The horizontal dotted line marks unity.}
     \label{fig:pk_LR_thermal}
\end{figure}

We have performed several supplementary simulations summarized in Table~\ref{tab:smallsims_supplementary}. Each of these simulations is designed to isolate specific factors that may influence the matter power spectrum suppression, such as resolution and feedback type (thermal versus kinetic). \textsc{SV-DF-LR} and \textsc{SV-Repos-LR} are low-resolution versions of the SV DF and repositioning runs, respectively. In \textsc{SV-DF-LR}/\textsc{SV-Repos-LR}, there are $2\times 500^3$ particles initially in the same volume as the SV runs listed in Table~\ref{tab:smallsims}, with a softening length of $3.3\,\mathrm{ckpc}/h$. \textsc{SV-Repos-T} differs from \textsc{SV-Repos} in that it only includes thermal feedback (i.e., even at low Eddington ratios, BHs in \textsc{SV-Repos-T} still inject feedback energy in the form of thermal energy instead of kinetic energy). \textsc{SV-Repos-Soffz1} is the same as \textsc{SV-Repos}, except that kinetic feedback is turned off from $z=1$ to $0$, which allows us to investigate the effects of kinetic feedback at early and late times separately.

\begin{table}
\centering
\caption{Supplementary SV simulations. A DMO counterpart, \textsc{SV-DMO-LR}, is also performed for the low-resolution runs.}
\begin{tabular}{lc}
\hline
\hline
Simulation & Description \\
\hline
\textsc{SV-DF-LR} & Lower resolution \\
\textsc{SV-Repos-LR} & Lower resolution \\
\textsc{SV-Repos-T} & Thermal feedback only \\
\multirow{2}{*}{\textsc{SV-Repos-Soffz1}} & Kinetic feedback turned off\\
&  since $z=1$\\
\hline
\end{tabular}
\label{tab:smallsims_supplementary}
\end{table}

\begin{figure*}[ht!]
     \centering
     \includegraphics[width=\textwidth]{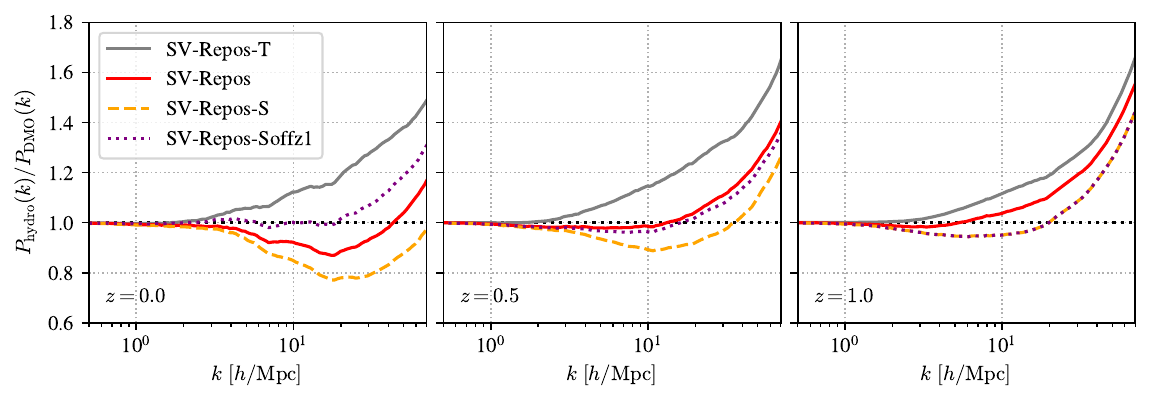}
     \caption{Comparison of $P(k)$ suppression among \textsc{SV-Repos-T} (solid gray), \textsc{SV-Repos-S} (dashed orange) and \textsc{SV-Repos-Soffz1} (dotted purple) at $z=0, 0.5$ and $z=1$ (left, middle and right panels, respectively). \textsc{SV-Repos} is also included for reference (solid red).}
     \label{fig:pk_Skz1}
\end{figure*}

Figure~\ref{fig:pk_LR_thermal} compares the matter power spectrum suppression among the simulations with different resolutions and feedback types at $z=0$. As expected, at low resolution (dashed--dotted), the discrepancy between the DF and repositioning runs is much less significant than at the original resolution (solid blue and red), because the larger softening length leads to more BH mergers and thus more efficient kinetic feedback in the DF run. Unlike the SV runs with the fiducial feedback model (in which kinetic feedback is allowed), the thermal-only run \textsc{SV-Repos-T} (solid gray) shows no suppression over the entire $k$ range, which verifies that kinetic feedback is essential for $P(k)$ suppression. In particular, \textsc{SV-Repos-T} shows enhanced power at $k\gtrsim 10\,h\,$Mpc$^{-1}$, by $\gtrsim 10\%$ compared with the fiducial runs (including \textsc{Astrid}). This implies that kinetic AGN feedback in \textsc{Astrid} has already damped matter clustering at these scales with respect to simulations without kinetic feedback, although the suppression is not significant enough to be detected in the ratio of $P_\mathrm{hydro}(k)/P_\mathrm{DMO}(k)$.

From Fig.~\ref{fig:pk_Skz1}, we see that the power ratio of \textsc{SV-Repos-S} is already significantly lower than that of \textsc{SV-Repos-T} at $z=1$ (right panel), indicating that the kinetic feedback in \textsc{SV-Repos-S} is effective at gas expulsion even at early times. As the simulation evolves, the power ratio of \textsc{SV-Repos-S} keeps decreasing while that of \textsc{SV-Repos-T} shows little change around $k\sim 10\,h\,$Mpc$^{-1}$. As for \textsc{SV-Repos-Soffz1}, its power ratio gradually approaches unity from $z=1$ to $z=0$ around $k= 10\,h\,$Mpc$^{-1}$, with its deviation from \textsc{SV-Repos-T} remaining quite significant at $z=0$. We note that, in \textsc{SV-Repos-S}, only $12\%$ of the total kinetic feedback energy is released prior to $z=1$ (see Fig.~\ref{fig:KE_density_BH}), but it accounts for almost half the total suppression at $z=0$ around $k\sim 10\,h\,$Mpc$^{-1}$ compared to \textsc{SV-Repos-T}. We also notice that, at $z=0.5$, the power ratio of \textsc{SV-Repos-Soffz1} is very close to that of \textsc{SV-Repos}, even though the total kinetic feedback energy released in \textsc{SV-Repos-Soffz1} is less than half of that in \textsc{SV-Repos}.
Early kinetic feedback is thus quite effective at suppressing matter clustering against cooling of baryons. This is likely because early halos have shallower potential wells and because the altered matter distribution can delay gravitational collapse.

\section{Halo matching method\label{app:halomatching}}

Here we introduce how we match halos between a pair of hydrodynamical and DMO simulations. For a certain halo in the hydrodynamical simulation, we first exclude all halos in the DMO counterpart that have masses
that differ from the hydrodynamical halo by $> 15\%$ and that are at a distance $> 100\,\mathrm{ckpc}/h$ from the hydrodynamical halo. The distance is measured between the potential minima of the two halos. Note that the mass and distance thresholds are used to shrink the search space and thus speed up the matching process, and they are not used as strict criteria to determine whether a halo pair is matched. We then calculate the number of shared DM particles between the hydrodynamical halo and each of the remaining DMO halos according to their particle IDs. A merit function is used to determine the best match, which is defined as
\begin{equation}
\mathcal{M}_{ij} = \frac{N_{ij}^2}{N_\mathrm{i}^\mathrm{hydro} N_\mathrm{j}^\mathrm{DMO}},
\end{equation}
where $N_{ij}$ is the number of shared DM particles between the $i$-th hydrodynamical halo and the $j$-th DMO halo, and $N_i^\mathrm{hydro}$ and $N_j^\mathrm{DMO}$ are the total number of DM particles in the $i$-th hydrodynamical halo and the $j$-th DMO halo, respectively. The DMO halo with the highest merit function value is considered as the best match for the hydrodynamical halo if the value is above 0.5. This particle-based matching technique follows methods commonly used in merger tree construction \citep[e.g.,][]{Srisawat2013}.

For each SV simulation presented in Fig.~\ref{fig:densityprofile_kfeedback}, we match halos using the above method to their DMO counterparts and keep only commonly matched halos in the four SV simulations for a fair comparison.

\section{Galaxy stellar mass function\label{app:GSMF}}

\begin{figure}[t!]
     \centering
     \includegraphics[width=\columnwidth]{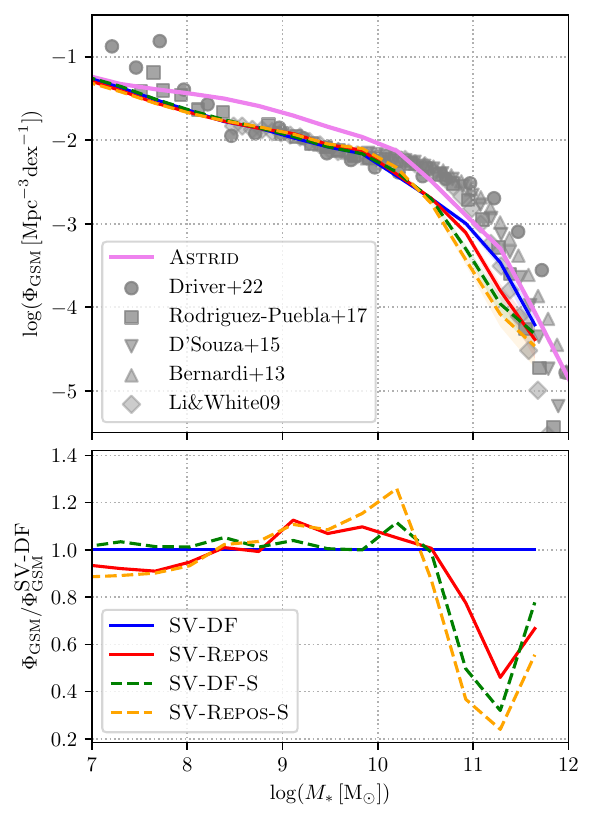}
     \caption{Top panel: GSMFs at $z=0$ measured from the \textsc{Astrid} and SV simulations (solid and dashed lines). Observational data from \cite{Driver2022,RodriguezPuebla2017,Li2009,Bernardi2013,DSouza2015} are plotted in gray markers. Bottom panel: Ratio of the GSMF of the SV runs to that of \textsc{SV-DF}.\label{fig:GSMF}}
\end{figure}

We examine the galaxy stellar mass function (GSMF), $\Phi_\mathrm{GSM} (M_*)$ (where $M_*$ refers to the galaxy stellar mass in $\mathrm{M}_\odot$), in the \textsc{Astrid} and SV simulations at $z=0$, as shown in the top panel of Fig.~\ref{fig:GSMF}. All galaxies (central and satellite) in the simulations are included in the GSMF measurement, and $M_*$ is defined as the stellar mass within twice the stellar half-mass radius. We also plot several observational constraints for comparison. The GSMF of \textsc{Astrid} (violet) is in good agreement with the observations, though it slightly overproduces galaxies at low masses ($M_\mathrm{*} \lesssim 10^{10}\,\mathrm{M}_\odot$) and underproduces galaxies at high masses. For more discussion of the GSMF in \textsc{Astrid}, we refer the readers to \cite{Zhou2026}.
At low and intermediate stellar masses, the GSMFs of the SV runs are similar to each other (while they are lower than that of \textsc{Astrid} due to the lower resolution). By contrast, at the high-mass end where kinetic AGN feedback is most effective, they show significant differences caused by the different BH dynamics and feedback models. 

To better illustrate the variations induced by different models, we plot the ratio of the GSMF of each SV run to that of \textsc{SV-DF} in the bottom panel of Fig.~\ref{fig:GSMF}. Compared to \textsc{SV-DF} (blue), \textsc{SV-Repos} (red) shows a lower GSMF at the high-mass end with a relative difference of up to $\sim 50\%$ at $M_* \approx 2\times 10^{11}\,\mathrm{M}_\odot$, due to the enhanced kinetic feedback. In the strong-feedback runs (dashed), the GSMF at the high-mass end is further suppressed, and we also see a more significant suppression in \textsc{SV-Repos-S} than in \textsc{SV-DF-S}.

\section{Gas fraction in halos\label{app:gasfraction}}

\begin{figure}[ht]
     \centering
     \includegraphics[width=\columnwidth]{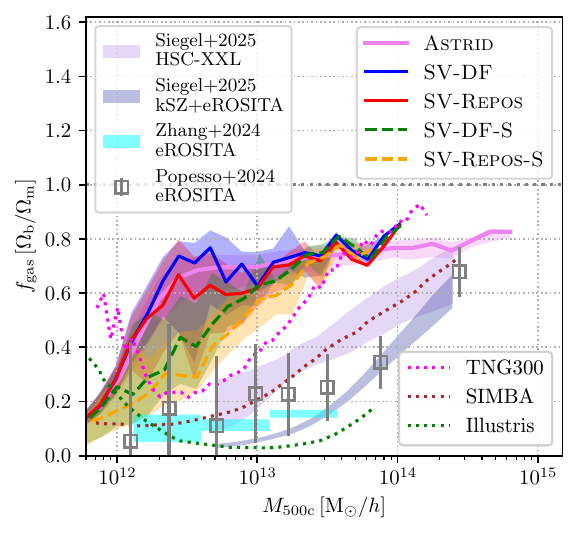}
     \caption{Gas mass fraction as a function of halo mass (normalized by the cosmic baryon fraction). Upper and lower bounds of shaded regions indicate the 16th and 84th percentiles. Line styles follow those in Fig.~\ref{fig:pk_kfeedback}. Constraints from the analyses of \cite{Siegel2025}, \cite{Zhang2024} and \cite{Popesso2024} are also shown for reference. Several other simulations are also plotted for comparison: IllustrisTNG, SIMBA \citep{Dave2019} and Illustris \citep[we use curves shown in][]{Popesso2024}.\label{fig:halo_gas_fraction}}
\end{figure}

In Fig.~\ref{fig:halo_gas_fraction}, we plot the mass fraction of hot gas, $f_\mathrm{gas} = M_\mathrm{gas}/M_\mathrm{500c}$, against halo mass for the hydrodynamical simulations at $z=0$. The gas mass is defined as the total mass of gas particles with temperatures $> 10^6\,$K within $R_\mathrm{500c}$. For comparison, we also plot observational constraints from \cite{Siegel2025} and \cite{Zhang2024} on X-ray (hot) gas mass fractions,\footnote{The observational data are not exactly at $z=0$, for example, the HSC-XXL samples are $z\sim 0.3$ clusters.} and results from selected other simulations (extracted from \citealt{Popesso2024}). Although \cite{Popesso2024} used only 10$^4$ halos from TNG300 to measure gas fractions, we have verified that using all halos in TNG300 gives broadly consistent results, despite their curve being biased high at the low-mass end.

Consistent with the results shown in Sec.~\ref{sec:clustering}, \textsc{SV-Repos} (red) has lower gas fractions than \textsc{SV-DF} (blue) \citep[the baryon fraction in groups correlates with the $P(k)$ suppression, see e.g.][]{Daalen2020}, especially at $M_\mathrm{500c} \lesssim 10^{13}\,\mathrm{M}_\odot/h$, due to the enhanced feedback. Likewise, in the strong-feedback case, \textsc{SV-Repos-S} shows more suppressed gas fractions than \textsc{SV-DF-S}.

\cite{Bahe2022} similarly found that instantaneous BH repositioning lowers the baryon fractions of massive halos relative to slower or no repositioning. However, because their comparison did not include an explicit physical DF model, it did not identify the repositioning-enhanced mergers as unphysically prompt or quantify their indirect effect on subsequent BH accretion and feedback.

Compared to other large-volume simulations (dotted), \textsc{Astrid} produces higher gas fractions across wide halo mass ranges, which is consistent with its milder kinetic feedback. While neither IllustrisTNG nor \textsc{Astrid} evacuates gas from halos as much as the observations suggest, SIMBA, which implements jet feedback, shows results in good agreement with the relatively low gas fractions from the constraints. In contrast, Illustris exhibits overly efficient gas expulsion \citep[see also][]{Popesso2024}. A simple order-of-magnitude estimate of the effect of using repositioning in the simulations suggests that a version of SIMBA with dynamical friction could have gas fractions that are too high, whereas a similar version of Illustris could match observations.

\bibliography{kfeedback}{}
\bibliographystyle{aasjournalv7}

\end{document}